\documentclass[12pt,aps]{revtex4}
\begin{document}

\title{Comment on the Quantum Brachistochrone Problem}

\author{C. M. Bender$^1$, D. C. Brody$^2$, H. F. Jones$^3$ and B. K. Meister$^4$}

\affiliation{$^1$Physics Department, Washington University, St. Louis, USA\\
$^2$Mathematics Department, Imperial College, London, UK\\
$^3$Physics Department, Imperial College, London, UK\\
$^4$Physics Department, Renmin University of China}

\begin{abstract}
In this brief comment we attempt to clarify the apparent
discrepancy between the papers \cite{CMB} and \cite{AM} on the
quantum brachistochrone, namely whether it is possible to use a
judicious mixture of Hermitian and non-Hermitian quantum mechanics
to evade the standard lower limit on the time taken for evolution
by a Hermitian Hamiltonian with given energy dispersion between
two given states.
\end{abstract}

\maketitle

We do not dispute the theorem of Ref.~\cite{AM}, which is a
generalization of Ref.~\cite{DB} and states that the limit can not
be evaded for unitary time development, but nonetheless maintain
that the effect identified in Ref.~\cite{CMB} is a real one. How,
then, is the theorem evaded? This can be explained in two ways.
First one could use the framework of standard quantum mechanics to
describe the entire process: preparation of the initial state,
time development under the influence of the non-Hermitian
Hamiltonian, and analysis of the final state. In that case,
clearly the time development is not unitary in the usual sense,
hence the possibility of an improved result. In this context the
importance of the $PT$ symmetry of the intermediate, non-Hermitian
Hamiltonian is that its energy eigenvalues are real, so that the
condition of the standard limit, namely that the energy dispersion
be fixed, makes sense.

The second way of explaining the improvement on the standard lower
limit is to use the alternative $CPT$ framework of quantum
mechanics for the time-development, whereby the intermediate
Hamiltonian is Hermitian, and hence the time-development unitary,
with respect to a new metric. The lower limit is now avoided
because the initial and final states, which we chose to be
orthogonal with respect to the standard metric, are no longer
orthogonal with respect to this new metric. However, in retrospect
it is to be admitted that we made too liberal use of the term
unitary in our original paper, because this unitarity was only
with respect to the $CPT$ operator defined for the intermediate
Hamiltonian. In the set-up where this intermediate Hamiltonian
represents a ``black box" in an otherwise Hermitian world, true
unitarity would have to refer to the $C$-operator for the entire
set-up, something we did not calculate.

\end{document}